\documentclass{article}
\usepackage{ijcai24}

\usepackage{times}
\usepackage{soul}
\usepackage{url}
\usepackage[hidelinks]{hyperref}
\usepackage[utf8]{inputenc}
\usepackage[small]{caption}
\usepackage{graphicx}
\usepackage{amsmath}
\usepackage{amsthm}
\usepackage{booktabs}
\usepackage{algorithm}
\usepackage{algorithmic}

\usepackage{multicol}
\usepackage{multirow}
\usepackage{subfigure}

\title{Interpersonal Relationship Analysis with Dyadic EEG \lyj{Signals} \\via Learning Spatial-Temporal Patterns}

\author{
Wenqi Ji$^1$\and
Fang Liu$^2$\and
Xinxin Du$^1$\and
Niqi Liu$^1$\and
Chao Zhou$^3$\and
Minjing Yu$^4$\and
Guozhen Zhao$^5$\thanks{Guozhen Zhao and Yongjin-Liu are the corresponding authors.}\and
Yong-jin Liu$^1$\footnotemark[1]
\affiliations
$^1$BNRist, Department of Computer Science and Technology, MOE-Key Laboratory of Pervasive Computing, Tsinghua University\\
$^2$State Key Laboratory of Media Convergence and Communication, Communication University of China\\
$^3$Institute of Software, Chinese Academy of Sciences\\
$^4$College of Intelligence and Computing, Tianjin University\\
$^5$CAS Key Laboratory of Behavioral Science, Institute of Psychology
\emails
\{jwq21, lnq22\}@mails.tsinghua.edu.cn,
\{duxx, liuyongjin\}@tsinghua.edu.cn,
fangliu@cuc.edu.cn,
zhouchao@iscas.ac.cn,
minjingyu@tju.edu.cn,
zhaogz@psych.ac.cn
}

\usepackage{color}
\newcommand{\ymj}{\textcolor{black}}
\newcommand{\lf}{\textcolor{black}}
\newcommand{\lnq}{\textcolor{black}}
\newcommand{\lyj}{\textcolor{black}}

\newcommand{\zc}{\textcolor{black}}

\begin{document}

\maketitle

\begin{abstract}
    Interpersonal relationship quality is pivotal in social and occupational contexts. Existing analysis of interpersonal relationships mostly rely on subjective self-reports, whereas objective quantification remains challenging. In this paper, we propose a novel social relationship analysis framework using spatio-temporal patterns derived from dyadic EEG signals, which can be applied to quantitatively measure team cooperation in corporate team building, and evaluate interpersonal dynamics between therapists and patients in psychiatric therapy. First, we constructed a dyadic-EEG dataset from 72 pairs of participants with two relationships (stranger or friend) when watching emotional videos simultaneously. Then we proposed a deep neural network on dyadic-subject EEG signals, in which we combine the dynamic graph convolutional neural network for characterizing the interpersonal relationships among the EEG channels and 1-dimension convolution for extracting the information from the time sequence. To obtain the feature vectors from two EEG recordings that \lyj{well} represent the relationship of two subjects, we \lyj{integrate} deep canonical correlation analysis and triplet loss for training the network. Experimental results show that the social relationship type (stranger or friend) between two individuals can be \lyj{effectively} identified through their EEG data.
\end{abstract}

\section{Introduction}

With the \lyj{development} of artificial intelligence \lyj{technologies}, EEG data has been extensively \lyj{explored} across various \lyj{research} areas and application scenarios, including emotion recognition \cite{al2017review}, depression or epilepsy detection \cite{2018Epileptic}, and \lyj{various implementations of brain-computer interface} \cite{2015eegbci}. In \lyj{deep-learning-based} EEG analysis, the majority of \lyj{existing} studies focus on tasks for \lnq{a} single person \cite{bashivan2016mental,zheng2017identifying}. However, \lyj{We note that} psychological investigations \cite{aftanas2004analysis} have found that there exists mutual influence in EEG data amongst individuals in a group, and their EEG data shows synchrony under certain tasks and stimuli. 

Inspired by this, \lyj{in this paper} we propose to \lyj{study} interpersonal relationships, which \lnq{are} categorized basically to communal sharing relationship\lnq{s} (e.g., friends, family members or couples) or equality matching relationship\lnq{s} (e.g., strangers) \cite{1994Categories}, by dyadic EEG data from two individuals. \ymj{The findings of our study can be used to objectively measure the quality of interpersonal relationships, for example, to guide the selection of team members and to circumvent the problem of cognitive bias that can arise when employees assess the level of rapport between them on their own.} \lyj{Our} work aims to \lyj{expand and} enlighten EEG-based research, shifting the focus from the neural variations of individuals to the interconnections in EEG patterns \zc{across individuals}.

Unlike the analysis of single-subject EEG data, the exploration of dyadic-subject EEG \lyj{data} presents significant disparities. Primarily, from a psychological perspective, the EEG data of paired participants reflect distinct mental phenomenon when faced with specific tasks compared to data from individual participants. Technologically, the analysis of dyadic-subject EEG data differs from the widely-researched individual EEG tasks like emotion recognition\cite{eegsurvey2019}, \lnq{requiring} a unique experimental design and procedure.
Moreover, for the construction and subsequent processing of datasets, studies \lyj{of} \zc{interpersonal interaction} require additional \lf{mechanisms} to analyze the relationship between the two participants. This multifaceted analysis extends beyond the scope of traditional single-subject EEG research and presents the unique challenges and complexities inherent to dyadic-subject EEG investigation.

\begin{figure*}[t]
    \centering
    \includegraphics[width=0.8\linewidth]{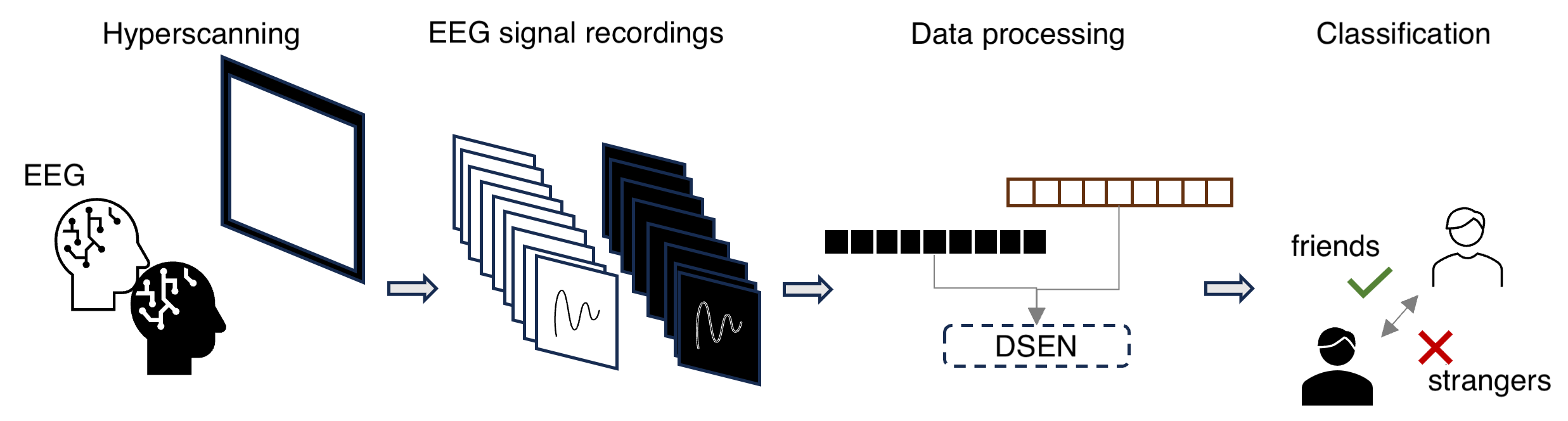}
    \caption{The procedure of exploring the relationships between two individuals. In our proposed framework, the relationship between the two participants (stranger or friend) can be determined based on their EEG recordings while watching \lyj{emotional} video clips.}
    \label{fig:flow}
\end{figure*}

\lyj{To address} the aforementioned challenges, our paper takes a pioneering step in the exploration of interpersonal relationships using dyadic-subject EEG \lyj{data} and AI methodologies. We have established a comprehensive process for analyzing dyadic-subject EEG data, covering problem definition, experimental design, dataset creation for dyadic-subject EEG, and the final algorithmic approach. As illustrated in Figure \ref{fig:flow}, we use hyperscanning, designed for simultaneously recording the brain activity of two or more individuals participating in the same cognitive activity \cite{2010hyperscan}, to construct our proposed dataset from the hyperscanned EEG recordings. Furthermore, we design a social relationship classification network tailored for dyadic-subject EEG data, which extracts spatial-temporal features from EEG sequence and, inspired by \cite{dcca2019}, combine the classification with \lyj{canonical correlation analysis} (CCA). To enhance the classification process, we have also integrated \lyj{a} triplet framework for enhancing the representations of the social relationship between the two subjects. 

Our contributions are summarized as follows:
\begin{itemize}
    \item We propose a novel EEG data analysis task to explore the social relationship between two individuals, and establish a pipeline for collecting and analyzing dyadic EEG data.
    \item We propose a network to model the temporal and spatial information of EEG \lyj{data}, where the EEG channels \lnq{are} modeled as nodes in \lnq{a} graph and processed with graph convolution operations while the time sequence is analyzed with 1-D convolution. For determining the relationships, we integrate the attention mechanism to fuse two types of features from dyadic EEG \lyj{data} and propose a training \lyj{scheme} with triplet loss and CCA loss.
    \item We construct a dyadic-subject EEG dataset for the analysis of social relationship between two individuals.
\end{itemize}

\section{Related works}
In this section, we first briefly delve into \lyj{methods} for EEG analysis 
based on deep learning. As mentioned previously, the majority of these techniques primarily concentrate on analyzing individual EEG data. \lyj{Contrastively, in this paper we} shift our focus to studies that specifically center on \zc{interpersonal} relations. 
It is noteworthy that some of these works \lyj{on multi-subjects} do not employ EEG data, while others refrain from leveraging AI-related technologies.

\subsection{Deep-Learning-based EEG Processing}
The exploration of emotion types or psychological states using EEG can be traced back to the 1980s \cite{1983earlyEEG}. Machine-learning-based methodologies for EEG analysis, including signal prediction and classification, \lnq{were} introduced in the early 2000s \cite{hazarika1997classification,dornhege2004boosting}. With the advance of computational power and the development of machine learning, including deep learning, an increasing number of EEG-based \lyj{methods and applications} have emerged. These studies leverage artificial intelligent technologies to accomplish tasks such as psychological diagnosis \cite{ho2023self}, and, most notably, emotion recognition \cite{al2017review}.

In order to explore the relationship between two individuals, discriminative features are required. Although most focused on \lnq{a} single subject, existing studies on EEG-based emotion \zc{measurement} have \lyj{spawned} effective methods for extracting effective EEG features and \lnq{have performed} well on tasks like emotion recognition. Regarding EEG-derived features, \cite{duan2013de,eegsurvey2019} indicated that features such as power spectral density (PSD) or differential entropy (DE) can be extracted. 
Further, \cite{ding2021interbrain} suggested that some transformations or combinations of the basic attributes of EEG signals, like amplitudes and phase, can serve as effective descriptors, especially for inter-brain EEG analyses. \lyj{Many} deep-learning-based algorithms have been proposed to tackle tasks related to emotions or brain activities using features from EEG. Among these studies, a subset based on \lyj{traditional} machine learning techniques, \lyj{e.g.,} the support vector machine (SVM) and random forest (RF), being the most prevalently utilized \cite{eegsurvey2019}. Within deep learning approaches, convolutional neural network (CNN) stands out 
given its capability to automatically learn spatial and temporal patterns from data. In the context of EEG-based CNN algorithms, for instance, EEG-Net \cite{eegnet} effectively learns information both temporally and \zc{inter-}electrodes in EEG data. 
As a representative of recursive network architectures, \cite{lstm2017} applied the LSTM to EEG and \lnq{achieved} good results in emotion recognition. Furthermore, considering that the electrodes \lf{and their relationships }in EEG can {be represented by} a graph, several graph convolutional networks (GCNs), have been proposed for EEG data \lf{analysis}. 
\lf{\cite{dgcnn2018} introduced a dynamic graph convolutional neural network (DGCNN) to simultaneously optimize network parameters and a weighted graph characterizing the strength of functional relation between each pair of two electrodes. SparseDGCNN~\cite{sparseDGCNN} modified DGCNN by imposing a sparseness constraint on the graph.}
Besides, 
\lf{other mainstream EEG analysis studies integrate various CNN and LSTM architectures to process EEG data, including deep belief network (DBN) \cite{zheng2015dbn}, variational pathway reasoning (VPR) \cite{zhang2020variational}, SST-EmotionNet~\cite{sstnet}, etc.} 
Moreover, there is a knowledge distillation framework \cite{liang2023teacher} and the SSLAPP method \cite{lee2022self} that employs attention modules for encoding the EEG signal and GANs for limited labeled data. 

\subsection{\lyj{Multi-Subjects Studies}}
Hyperscanning \cite{hyperscanning} is a technique that involves the simultaneous recording of brain activity from two or more individuals engaged in the same task or social interaction. It is a widely used experimental paradigm for multi-subject studies.

To \lnq{analyze} synchronization and mutual psychological influence on individuals within \lnq{a} group, existing studies often involve physiological signals other than EEG or \lnq{do} not employ AI-based methods. 
\cite{gamliel2021inter} investigated paired subjects performing tasks together using functional near-infrared spectroscopy (fNIRS), and their study revealed that certain brain regions, specifically the inferior frontal gyrus (IFG), played a pivotal role during the task. Similarly,  \cite{pan2017cooperation} also cope up with fNIRS 
to investigate dyads consisting of lovers, friends, and strangers. 
They asked subjects to perform a cooperation task and recorded the brain activities in \lnq{the} right frontoparietal regions.
With EEG data, \cite{ding2021interbrain} conducted a comparative analysis to identify EEG features suitable for inter-brain exploration. 
These studies validated the existence of brain synchronization during multi-subject activities, thereby enabling the exploration of relationships between two subjects through their neural signal data.
In this paper, \lyj{we investigate a novel task} to determine the relationship type of the two participants using dyadic-subject EEG data.

\section{Methods}

\begin{figure*}[t]
    \centering
    \includegraphics[width=0.85\linewidth]{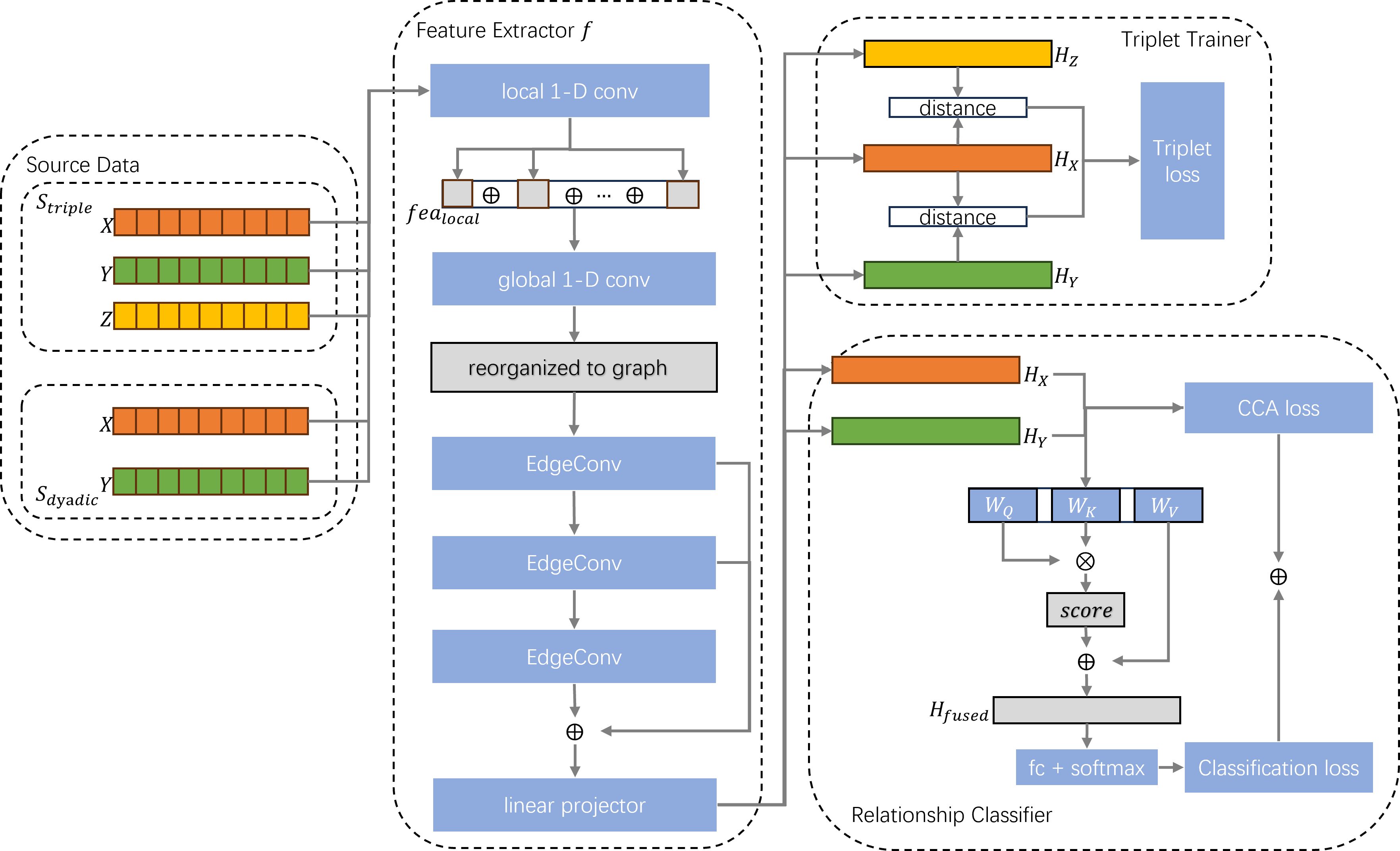}
    \caption{The framework of our proposed Dyadic-Subject EEG Network (DSEN). DSEN takes dyadic-subject EEG data as input to extract temporal and spatial features with the feature extractor $f$. The features are fused by an attention-based module to form a cohesive representation for classification. The feature extractor $f$ is trained individually via a triplet loss. The fused features are utilized by a classifier, which is trained using a combined CCA loss and classification loss.}
    \label{fig:framework}
\end{figure*}

\subsection{Dyadic-Subject EEG Dataset}
\subsubsection{EEG Data Collection}
\label{experiments}
 In our experiment, EEG signals were recorded using the NuAmps$^\text{TM}$ 40-channel unipolar system (Compumedics Neuroscan, USA) with \lnq{a} sampling frequency of 1000Hz. Two reference electrodes were attached to the mastoid areas on both sides of the subjects. The impedance of all electrodes was maintained below $5 k\Omega$. A low-pass filter was set to 60Hz.
 
 Seventy-two pairs of healthy participants, aged between 19 and 30, were recruited from universities and participated in the experiment. Among all these pairs of participants, 18 pairs were strangers to each other, while the remaining 54 pairs were friends or couples. After the EEG recording system was fitted to the subjects, the pair\lf{s} of subjects were asked to sit still for 8 minutes to EEG signals are collected from their resting state. The subjects were then asked to sit upright in a chair and to clear their minds. During the collection of EEG data, the subjects were seated half a meter apart and were not allowed to communicate with each other. 
 
\begin{figure}[h]
    \centering
    \includegraphics[width=0.9\linewidth]{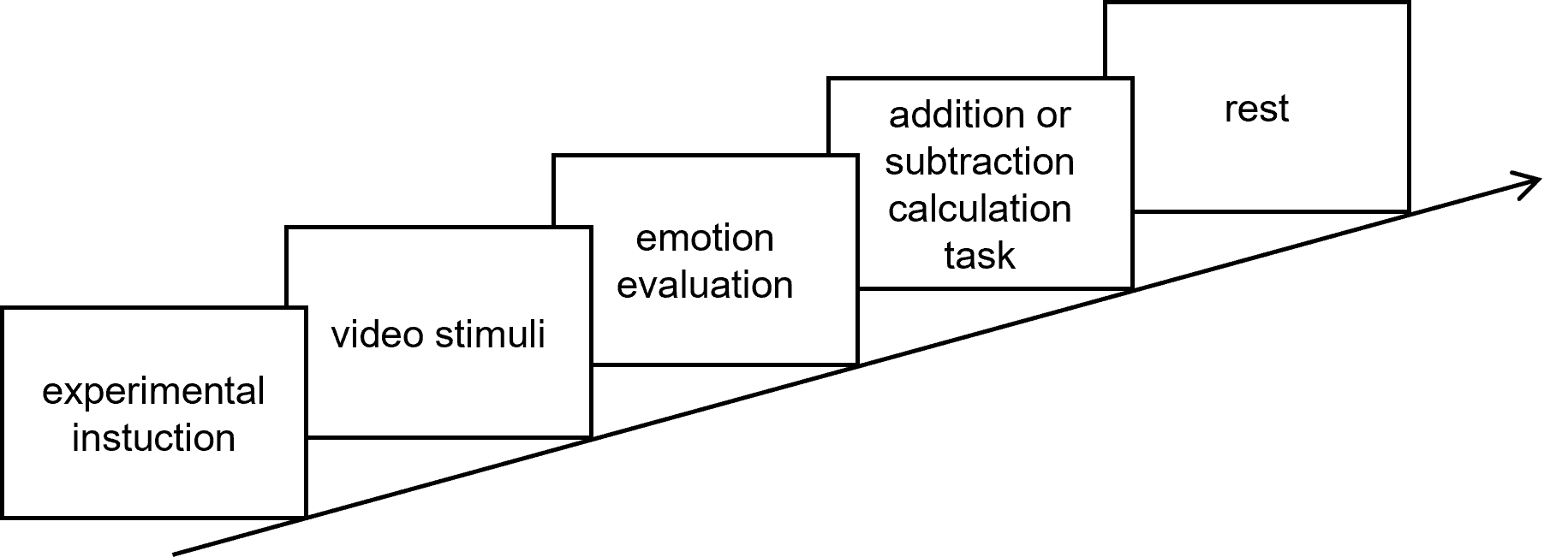}
    \caption{The trial procedure of our dyadic-subject EEG data collection. }
    \label{fig:trial}
\end{figure}

 For main experimental tasks, the two subjects \lnq{who} participated in the hyperscanning experiments were asked to watch nine video clips. Given that positive emotion exhibits more synchronization than negative emotions in interpersonal relationships \cite{2015Positive}, video clips labeled with positive emotions were selected from the Chinese emotional film database proposed by \cite{videodb} for the experiments. Moreover, to provide a richer emotional perspective, nine video clips representing the positive emotions of friendship, awe, kinship, respect, gratitude, humor, longing, pride, and love were chosen. The duration of the video clips ranged from 160 seconds to 296 seconds, with an average length of 216 seconds. After watching each \lyj{emotional} video clip, subjects were required to perform a distractive task to neutralize the emotional effect of the previously \lyj{watched} video clip, and were given a mandatory half-minute break before the next video clip was presented. The above process is repeated until all 9 video materials have been played.
 The procedure of a trial is shown in Figure \ref{fig:trial}. 
 
 All procedures of the study were reviewed and approved by the Ethics Committee of Tsinghua University. Before the experiment, all the participants were explained well about the details of the experiment and signed the informed consent.  

\subsubsection{Preprocessing}
\label{dataset}
Through the aforementioned experiment, we collected our dyadic-subject EEG data that includes hyperscanned EEG signal recordings of 144 participants as 72 pairs\footnote{All EEG data comprise a total of 5,688 data samples after window-based segmentation and concatenation.}, \lyj{who were} co-watching nine \lyj{emotional} video clips. For subsequent processing and analysis, we downsampled the data to 200Hz, applied a digital filter from 1-45Hz, and then re-referenced using the average of the left and right mastoids. Additionally, on this basis, ocular artifacts were removed using independent component analysis (ICA). In order to evaluate our proposed DSEN framework (Section \ref{subsec:DSEN}), we segmented the EEG signal recordings of subjects watching each video clip into 2-second windows. Then, we concatenated the 2-second segments from all nine video clips to form a data sample, thereby establishing our dyadic-subject EEG dataset. 

\subsection{Preliminary Study}
Currently widely-used EEG datasets, such as DEAP \cite{deap}, SEED \cite{zheng2015dbn}, and MPED \cite{mped}, \lyj{only contain} data of individual \lyj{subjects} and are not suitable for analyzing the relationship between two subjects. \lyj{We} set up an experiment to collect dyadic-subject EEG data for exploring the social relationship between two individuals. Sec.~\ref{experiments} provides the detailed construction process of our dyadic-subject EEG dataset.
Then we conduct \lnq{a} specific statistical test on the \lyj{dyadic-subject EEG data} to validate the correlation between the \lf{participants} relationships and the EEG signals.

\subsubsection{Notations}
In our dataset, each data sample represents EEG data of a pair of two subjects, denoted by $(X,Y)$.
As input for the subsequent classification and analyses \lnq{tasks}, we derive the amplitude sequences denoted by $(A_X,A_Y)$ from $(X,Y)$. Let $n$ representing the number of \lyj{emotional} video clips that the two subjects co-watched, each input sample $(a_x, a_y)= \{ (a_x, a_y)^1, (a_x, a_y)^2, \cdots, (a_x, a_y)^n \} \in (A_X, A_Y)$ contains $n$ EEG segments from the two subjects co-watching all videos clips with different labels of emotions. Specifically, as applied with the same sliding window to segment the EEG recordings, each \lyj{sample segment} $(a_x, a_y)^i \in (a_x, a_y)$ has an equal length of time steps, ensuring that different emotional stimuli receive equal consideration.  

\subsubsection{Statistic Analysis}
\label{statanalysis}
In our proposed innovative analysis process, we first verify the data distribution, and then conduct preliminary feature selection and processing. 

We employ a Shapiro-Wilk (S-W) test to verify that the EEG data we collected conforms to a normal distribution. Consequently, we can use the t-test method on the band-pass filtered EEG data to verify the influence of the relationship and gender pairing of the dyadic subjects.
This initial analysis serves to explore the correlation between the EEG data of the two subjects \lyj{with respect to} their relationships. Additionally, these procedures also effectively alleviate the potential influence of gender on synchrony differences. For S-W test 
and t-tests, the instantaneous amplitude and phase samples are generated from the analytic signals $X_{analytic}(t)=X(t)+iX_H(t), Y_{analytic}(t)=Y(t)+iY_H(t)$, obtained from Hilbert transform \cite{le2001comparison}. The amplitude and phase samples are derived as:
\begin{equation}
\begin{split}
    A_X(t) &= \sqrt{X(t)^2+X_H(t)^2}, A_Y(t) = \sqrt{Y(t)^2+Y_H(t)^2} \\
    \phi_X(t) &= tan^{-1}\frac{X_H(t)}{X(t)}, \phi_Y(t)=tan^{-1}\frac{Y_H(t)}{Y(t)}
\end{split}
    \label{eq:amp}
\end{equation}
Subsequently, the inter-subject correlation $ICS_{XY}$ and the phase locking value $PLV_{XY}$ are derived by:
\begin{equation}
\begin{split}
    ISC_{XY}&=\frac{cov(A_X(t), A_Y(t))}{\sigma A_X(t) \sigma A_Y(t)} \\
    PLV_{XY}&=|\frac{1}{N}\sum_{n=1}^Ne^{i(\phi_X(tn)-\phi_Y(tn)}|
\end{split}
\end{equation}
With the $ISC_{XY}$ and $PLV_{XY}$, we applied \lnq{a} t-test for both the relationship status of the pair (whether they know each other or not) and the gender pairing (whether they are of the same gender or different genders). We also compared the significance of dyadic relationship differences across various frequency bands, namely $\theta$ (4-7Hz), $\alpha$ (8-12Hz), $\beta$ (13-29Hz), and $\gamma$ (30-45Hz). This 
analyses enabled us to pre-select the optimal frequency band combination, which are $\beta$ and $\gamma$ bands, for subsequent processing. 

\subsubsection{T-test Results}
We used the ISC and PLV features in different frequency bands as dependent variables. The type of relationship between them \lnq{was} used as \lnq{an} independent variable for the t-test. Specifically, the types of relationships were divided into freindss and strangers. These categorizations facilitated the exploration of potential influences on the dependent variables. 

\begin{table}[t]
    \centering
    \begin{tabular}{l|l|c|c}
    \hline
    IV & feature & statistic & p-value \\
    \hline
    \multirow{8}{*}{relation} & $PLV_{\theta}$ & -0.400 & 0.691 \\
     & $PLV_{\alpha}$ & 0.587 & 0.559 \\
     & $PLV_{\beta}$ & 0.282 & 0.779 \\
     & $PLV_{\gamma}$ & -1.080 & 0.284 \\
     & $ISC_{\theta}$ & 0.735 & 0.465 \\
     & $ISC_{\alpha}$ & -0.250 & 0.803 \\
     & \textbf{$ISC_{\beta}$} & \textbf{1.787} & \textbf{0.078} \\
     & \textbf{$ISC_{\gamma}$} & \textbf{2.287} & \textbf{0.025} \\
    \hline
    \end{tabular}
    \caption{t-test results with $ISC$ and $PLV$ from dyadic-subject EEG data as dependent variables and relationships of them as independent variables.}
    \label{tab:ttest}
\end{table}
The significance level was set to 0.05. Table \ref{tab:ttest} demonstrates that the EEG synchrony features $ISC$ for dyadic subjects show significant differences across different relationship types in $\beta$ frequency band and marginally significant in $\gamma$ band. Our experimental results show that synchronized dyadic EEG activities in the $\beta$ and $\gamma$ frequency bands are effective in distinguishing the relationships between two individuals. This is also supported by some existing findings that the $\beta$ or $\gamma$ frequency bands play a more significant role in regression analysis for inter-brain emotion prediction \cite{ding2021interbrain}. 
These \lnq{discoveries} in the preliminary study provided theoretical foundations for our algorithmic design of predicting relationships based on EEG data and our pre-selected EEG features. The results of the t-test for gender types are shown in the supplementary. 



\subsection{Feature Extraction}
In order to extract efficient features from the raw EEG data i.e., $(A_X,A_Y$), we propose a sequential processing framework that mainly contains two modules (see the feature extractor block in Figure~\ref{fig:framework}):
1) a CNN module to extract the temporal information from the EEG amplitude sequences, and 2) a DGCNN module to identify the relationships among EEG electrodes.

Considering that our input includes EEG data segments of dyadic subjects co-watching nine different video clips \lyj{(see Sec.~\ref{dataset} for details)}, a grouped 1-D convolutional layer is used to extract the local temporal features $\{fea_{local}^1,\cdots,fea_{local}^9\}$ from each segment of $a_x$ and $a_y$, and the outputs are concatenated into $F = fea_{local}^1\oplus \cdots \oplus fea_{local}^9$ as local temporal feature outputs and then \lyj{are} used as the input of following global 1-D convolutional layer. The second 1-D convolutional layer is applied to obtain the overall temporal features $T_x, T_y$ from $F_x, F_y$. Within the CNN module described above, we apply convolutions separately for each channel of EEG. 

\lyj{Although} the CNN-based layers \lnq{ensure} that our temporal information extraction module can effectively \lf{learn the time-series features} in EEG data, they do not account for inter-electrode relationship information. 
Consequently, a DGCNN is employed for feature extraction from the EEG electrodes. We set the adjacency matrix $A$, constructed as a fully-connected graph, to represent the connectivity between each pair of \lnq{vertices} in the graph. Taking EEG electrode channels as vertices (denoted by $ch$), the $t = ch \times fea \in \mathcal{T}$ could be reorganized to form the vertex set $V$, and then we set the input graph $G = \{ V, A \}$. Inspired by \cite{dgcnn}, we introduce three EdgeConv blocks. Specifically, for a vertex $v_i$ in $V$ and one of its adjacent vertices $v_i' \in V'$, a transformed feature $h_i$ of $v_i$ is obtained from EdgeConv by 
\begin{equation}
    h_i=\phi_e({fea}_i'-{fea}_i)
\end{equation}
where $\phi_e$ denotes the shared multi-layer perceptron (sMLP) of the $e^{th}$ EdgeConv block. Taking $H_{pooled}^{1\sim 3}$ as global max pooled vertex features from three EdgeConv blocks, the feature $H$ extracted by DGCNN, which is also the output feature of our proposed extractor, is obtained from $\varphi(H_{pooled}^1 \oplus H_{pooled}^2 \oplus H_{pooled}^3)$, where $\varphi$ is linear transformations. 

\subsection{Dyadic-Subject EEG Network \lyj{(DSEN)}}
\label{subsec:DSEN}
As shown in Fig. \ref{fig:framework}, \lyj{our} proposed dyadic-subject EEG network (DSEN) \lyj{consists of} a triplet network and a classifier, \lyj{in} which we first \lyj{package} our source data from a triplet set and a dyadic set as $S=\{S_{tri}, S_{dual}\}$. The triplet source data is set as $S_{tri}=\{X, Y, Z\}$, where $X$ and $Y$ represent dyadic subjects (selected for $S_{tri}$ with ground truth of relationship types), while $Z$ is a subject selected randomly from the other participants, being a stranger either to $X$ or $Y$. The $S_{dual}$ is formed as $\{X, Y, L\}$, where $L$ is the set of labels with values of 0 or 1 indicating that the dyadic subject $X$ and $Y$ are strangers or freinds, respectively. \lyj{To update $\hat{\theta}_f$ in training,} the weight of $\textit{f}$ via triplet loss, the distance of cosine similarity between anchor ($X$) and positive ($Y$) or negative ($Z$) is calculated on the output features $\{ H_X, H_Y, H_Z\}$ from the extractor. \lyj{Then} $\hat{\theta}_f$ \lyj{is} updated with back propagation of $\mathcal{L}_{triplet}$. To train the classifier, we fuse the two features $\{H_X, H_Y\}$ extracted from $S_{dual}$. For the feature fusion, we integrated the attention mechanism since we perceive $H_X$ and $H_Y$ as sequences of embeddings, and interactions and relationships between them could be captured. To obtain the matrices of query ($Q$), key ($K$) and value ($V$) for both of $H_X$ and $H_Y$, each of the two features \lnq{is} transformed linearly using learnable matrices $W_Q$, $W_K$ and $W_V$ by:
\begin{equation}
    Q_{X,Y},K_{X,Y},V_{X,Y}=W_{Q,K,V} \times H_{X,Y}.
    \label{eq:matrixQKV}
\end{equation}
Then the attention score $S$ between $H_X$ and $H_Y$, and the attention-based fusion feature $H_{fused}$ \lyj{are} computed by:
\begin{equation}
\begin{split}
    S_{X,Y} &= \frac{Q_{X,Y} \times K_{Y,X}^T}{c} \\
    H_{fused} &= (softmax(S_X)\times V_Y) \oplus (softmax(S_Y)\times V_X),
    \label{eq:fusedfeature}
\end{split}
\end{equation}
where $c$ is a constant \lyj{scale} factor that is set to $128^{0.5}$ in our experiments. Subsequently, $H_{fused}$ is used to generate the prediction for \lyj{stranger} and \lyj{freind} relationships of the dyadic subjects by passing through fully connected layers. To update both $\hat{\theta}_f$ and the weight of the classifier $\hat{\theta}_c$, we propose a combined loss \lyj{defined} as:
\begin{equation}
    \mathcal{L}_{Combined}=\alpha \cdot \mathcal{L}_{Classification} + \beta \cdot \mathcal{L}_{CCA}
    \label{eq:combinedloss}
\end{equation}
where $\mathcal{L}_{Classification}$ denotes the cross-entropy loss, and $\alpha,\beta$ are the combined factors, which are both set to 1 in our experiments. Algorithm \ref{alg:DSEN} illustrates the complete training procedure of our proposed DSEN framework. \lyj{Details} of the triplet loss and the CCA loss used in our \lyj{method} is \lyj{presented in the following section.}

\begin{algorithm}[t]
\caption{training of DSEN}
\label{alg:DSEN}
\textbf{Input}: data set $S_{tri}, S_{dual}$\\
\textbf{Output}: parameters $\hat{\theta}_f$, $\hat{\theta}_c$
\begin{algorithmic}
\STATE calculate the instantaneous amplitude set $\{A_X, A_Y, A_Z\}$ by $\{X, Y, Z\}$ from $S_{tri}, S_{dual}$ via Equation \ref{eq:amp}
\WHILE{not converge and not reach the max number of iterations}
\STATE $fea_{local \cdot X,Y,Z}=CNN1D_{local}(A_{X,Y,Z})$
\STATE $T_{X,Y,Z}=CNN1D_{global}(fea_{local \cdot X,Y,Z})$
\STATE $H_{X,Y,Z}^1=EdgeConv_1(T_{X,Y,Z})$
\STATE $H_{X,Y,Z}^2=EdgeConv_2(H_{X,Y,Z}^1)$
\STATE $H_{X,Y,Z}^3=EdgeConv_3(H_{X,Y,Z}^2)$
\STATE $H_{X,Y,Z}=H_{X,Y,Z}^1 \oplus H_{X,Y,Z}^2 \oplus H_{X,Y,Z}^3$
\STATE Use $H_{X,Y,Z}$ of $X,Y,Z \in S_{tri}$
\STATE compute $\mathcal{L}_{triplet}$ by Eq. \ref{eq:distcos} and \ref{eq:triloss}
\STATE update $\hat{\theta}_f$ with $\frac{\partial{\mathcal{L}_{triplet}}}{\partial{\theta_f}}$
\STATE use $H_{X,Y}$ of $X,Y \in S_{dual}$
\STATE compute covariance $R_{X,Y,XY}$
\STATE compute $\mathcal{L}_{CCA}$ by Eq. \ref{eq:matrixE} and \ref{eq:ccaloss}
\STATE get $H_{fused}$ by Eq. \ref{eq:matrixQKV} and \ref{eq:fusedfeature}
\STATE $preds=Classifier(H_{fused})$
\STATE $\mathcal{L}_{Classification}=Classification\textit{loss}(preds, labels)$
\STATE compute $\mathcal{L}_{Combined}$ by Eq. \ref{eq:combinedloss}
\STATE update $\hat{\theta}_f$ and $\hat{\theta}_c$ with $\frac{\partial{\mathcal{L}_{Combined}}}{\partial{\theta_f}}$ and $\frac{\partial{\mathcal{L}_{Combined}}}{\partial{\theta_c}}$
\ENDWHILE
\STATE \textbf{return} $(\hat{\theta}_f, \hat{\theta}_c)$
\end{algorithmic}
\end{algorithm}

\subsection{Loss Function}

\subsubsection{CCA Loss}
Inspired by \cite{dcca2019} that integrates canonical correlation analysis (CCA) with deep neural networks for multimodal language analysis, we introduced \lyj{the} CCA loss to learn the correlation between the dyadic features extracted by the proposed feature extractor. The features $H_X, H_Y$ \lyj{are} normalized through mean centering, which ensures zero mean for each feature, and facilitates further covariance computation. With $H_X^{adapted}, H_Y^{adapted}$, we \lyj{then} calculate the covariance matrices of the two features $R_X ,R_Y $ and the cross-covariance $R_{XY}$. Our objective with CCA loss is to maximize the correlation between the features extracted by the feature extractor $f$, as shown in the equation:
\begin{equation}
    \max_{\textit{f}} \textit{corr} (\textit{f}(A_X),   \textit{f}(A_Y))
    \label{dccaf}
\end{equation}
the matrix $E$ corresponding to canonical correlation coefficients between the two features is derived \lyj{as}:
\begin{equation}
    E = R_X^{-\frac{1}{2}} R_{XY} R_Y^{-\frac{1}{2}}. 
    \label{eq:matrixE}
\end{equation}
To update the parameters of $\textit{f}$, \lyj{in} (\ref{dccaf}), CCA loss is defined as:
\begin{equation}
    \mathcal{L}_{CCA}=-trace(E^TE)^{\frac{1}{2}}=-\sum s,
    \label{eq:ccaloss}
\end{equation}
where $s$ are the singular values of $E$ and derived from singular value decomposition of $E$.
\subsubsection{Triplet Loss}
To further train the $\textit{f}$ to discriminate the relationship type between two subjects, we introduce the triplet loss as an additional criterion that helps $\textit{f}$ to learn more distinctions by comparing a pair of similar samples against a dissimilar sample. Given an $anchor$ sample, a $positive$ sample, and a $negative$ sample, we \lyj{formulate the} computation of two samples based on cosine similarity as:
\begin{equation}
    \textit{dist}_{p/n}=1.0-\frac{anchor \cdot positive/negative}{\left \| anchor \right \|^2 \times \left \| positive/negative \right \|^2}
    \label{eq:distcos}
\end{equation}
Then the triplet loss \lyj{is defined as}:
\begin{equation}
    \mathcal{L}_{triplet}=ReLU(\textit{dist}_p - \textit{dist}_n + margin).
    \label{eq:triloss}
\end{equation}

\section{Experiments}

\subsection{Implementation}
To validate the efficiency of our proposed DSEN on discriminating the social relationship type between two subjects, we compared six state-of-the-art methods for processing EEG data on our proposed dyadic-subject EEG dataset. Considering our dyadic data and the task of classifying the relationship, rather than emotion recognition as many of these methods were originally designed for, we implemented these models to extract features from EEG data of each individual separately. We then simply concatenated the features extracted from the EEG of both subjects as feature fusion and fed them through two linear layers for classification. All methods were trained and tested on our dyadic-subject EEG dataset. \lyj{Noting} that our dataset is imbalanced, with 18 pairs of subjects who are strangers and 54 pairs who are \lnq{freinds} to each other, and in order to verify the subject-independence of our model, we partitioned our training and test sets as follows: we selected 15 pairs of subjects from both the strangers and freinds relationships. The data samples from these 30 pairs, segmented using a 2-second window and then concatenated across all 9 video segments, were used to form the training set. The data samples from all remaining subject pairs constituted the test set. 

\begin{table}[h]
    \centering
    \begin{tabular}{l|c|c}
        \hline
        method & accuracy & F1 score \\
        \hline
        SVM & 0.65 & 0.78 \\
        DBN & 0.72 & 0.83 \\
        DGCNN & 0.61 & 0.75 \\
        LSTM & 0.58 & 0.72\\
        EEGNet & 0.67 & 0.79 \\
        SST-EmotionNet & 0.64 & 0.76 \\
        SSLAPP & 0.77 & 0.86 \\
        \textbf{DSEN (ours)} & \textbf{0.86} & \textbf{0.92}
         \\     \hline
    \end{tabular}
    \caption{Results \lyj{of} different methods applied on our proposed dyadic-subject EEG dataset.}
    \label{tab:expresult}
\end{table}

\subsection{Classification Results}
The methods we used \lyj{for} comparison are SVM \cite{1999svm}, DGCNN \cite{dgcnn2018}, LSTM \cite{lstm2017}, EEGNet \cite{eegnet}, DBN \cite{zheng2015dbn}, SST-EmotionNet \cite{sstnet} and SSLAPP \cite{lee2022self}. Considering that there are currently no deep learning methods designed to handle dyadic EEG data, in our experiments, we selected some representative methods for EEG-based emotion recognition or sleeping stage classification. These methods were utilized as encoders for EEG data. Subsequently, the EEG features extracted from two individuals were concatenated, and \lnq{an} MLP classifier was employed to classify their relationship. In particular, as the traditional SVM model \lyj{requires} the input features to be a one-dimensional vector. Therefore, for the SVM classifier, preprocessing of the data is \lyj{done} that we first flatten each data sample from \lnq{the} dyadic EEG dataset and then compute the Pearson correlation between the dyadic samples over the entire time step range to serve as the artificial $ISC$ features for the SVM. 
The accuracy and F1 score results are reported in \lyj{Table} \ref{tab:expresult}. Compared to the \lyj{baseline} methods, including CNN-based, GCN-based, RNN-based, and traditional machine learning approach \lyj{SVM}, our method achieved a \lyj{significant} improvement of at least 14 percent in classification accuracy for determining the relationship type between the two subjects, reaching over 85 percent. \lyj{meanwhile}, in terms of the F1 score, our proposed DSEN also outperforms other methods \lnq{by} at least 0.09. The results indicate that the mechanism introduced in our model is more suitable for analyzing dyadic-subject EEG data. Compared to \lyj{those} algorithms designed for tasks like emotion recognition for individuals, our \lyj{method} can predict the relationship between two participants more effectively.

\begin{figure}[h]
\begin{minipage}[t]{0.45\linewidth}
    \centering
    \subfigure[]{
    \includegraphics[height=3.5cm,width=3.5cm]{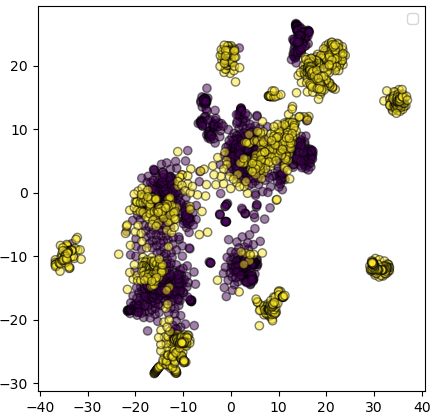}
    }
    \end{minipage}
    \begin{minipage}[t]{0.45\linewidth}
    \centering
    \subfigure[]{
    \includegraphics[height=3.5cm,width=3.5cm]{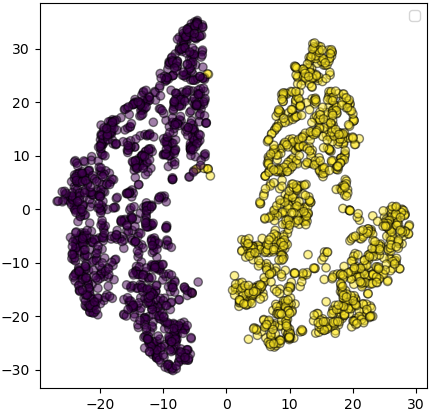}
    }
    \end{minipage}
    \caption{(a) is the distribution plot of the unprocessed dyadic EEG data in a two-dimensional plane after dimension reduction \lyj{by t-SNE}. (b) is the distribution plot of extracted and fused features from the dyadic EEG data in the plane after \lyj{dimension reduction by t-SNE}.}
    \label{fig:vis}
\end{figure}
\subsection{Feature Discrimination Comparison}
To validate the feature extraction capability of our proposed model for dyadic-subject EEG \lyj{data}, we conducted an experiment to visually compare the features extracted by our feature extractor and fused using our methods, with the unprocessed EEG amplitude data. These features were assessed based on their ability to discriminate different classes of data samples in our dyadic-subject EEG dataset. In the experiment, we employed t-distributed stochastic neighbor embedding (t-SNE) to reduce the dimension of the high-dimensional features. The features were then visualized in a 2D coordinate space. As illustrated in \lyj{Fig.} \ref{fig:vis}, it is \lyj{obvious} that the feature vectors extracted by our model from the dyadic-subject EEG can effectively distinguish between different social relationship types. 

\subsection{Ablation Study}
In order to validate the efficiency of the specific methods introduced in our framework, we conducted a series of ablation studies. We compared the performance of our full model against various ablated versions, including those without the combined CCA loss, without the triplet loss, and versions where the features extracted from the feature extractor were directly concatenated instead of using an attention mechanism for fusion. The removal of CCA loss showed a notable decrease in performance, indicating its vital role in enhancing the correlation learning between the data of participants. Similarly, excluding the triplet loss led to reduced accuracy, showing its importance in optimizing the feature space for better discrimination and classification. The absence of an attention mechanism resulted in a lower F1 score, illustrating its efficacy in weighting and fusing features from the dyadic participants. Additionally, we also evaluated a version that employs raw EEG data without transforming \lnq{it} to compute the instantaneous amplitude of the EEG waveform. This approach significantly underperformed, demonstrating the necessity of data transformation for effective feature extraction and prediction. The outcomes of these ablation studies, presented in Table \ref{tab:ablation}, \lyj{indicate} that the proposed enhancements in our framework significantly improve the prediction of the relationship between the two participants.
\begin{table}[h]
    \centering
    \begin{tabular}{l|c|c}
        \hline
        modifications & accuracy & F1 score \\
        \hline
        without CCA loss & 0.81 & 0.89 \\
        without triplet & 0.78 & 0.87 \\ 
        without attention & 0.81 & 0.89 \\
        raw EEG data & 0.52 & 0.66 \\
        \textbf{full model} & \textbf{0.86} & \textbf{0.92}
         \\     \hline
    \end{tabular}
    \caption{Results of ablation studies.}
    \label{tab:ablation}
\end{table}

\section{Conclusion}
In this \lyj{paper}, we introduce a novel research \lyj{task} that aims to explore the relationship between two individuals based on their EEG data. To \lyj{accomplish this task}, we delineated a comprehensive \lyj{strategy} encompassing experiment design, dataset construction, the preliminary statistical analyses that \lnq{indicate} the connection between the social relationship types and the dyadic EEG signals, an initial feature selection, and a DSEN framework designed for processing dyadic-subject EEG data. Our proposed classification algorithm, tailored for dyadic EEG \lyj{data}, incorporates advanced mechanisms such as CCA loss, triplet training, and attention-based feature fusion. By comparing our \lyj{method with baseline} algorithms designed for individual EEG emotion recognition, our results on the proposed dyadic-subject EEG dataset \lnq{show} its effectiveness \lnq{in} predicting interpersonal relationship \zc{tpyes}. Our ablation studies further \lyj{emphasize} the importance of each component we introduced. 


In conclusion, our results demonstrate the potential of our framework in effectively utilizing dyadic EEG data to predict relationships. The introduced mechanisms and \lyj{strategy} have set a benchmark in this \lyj{novel research} area, drawing more attention to the exploration of interpersonal relationship types based on the synchrony of inter-brain within a group using EEG. 

\bibliographystyle{named}
\bibliography{ijcai24}

\clearpage

\section{Supplementary Materials}

\subsection{Detailed Implementation of Dyadic-Subject EEG Network (DSEN)}
In this section, we provide a comprehensive implementation of our proposed DSEN framework, detailing the components of each module and the parameter configurations. Given the descriptions in our paper, the dyadic-subject EEG data comprises 30 electrode channels and the sampling rate has been downsampled to 200Hz. Additionally, we utilize a 2-second sliding window to obtain segments from the 9 EEG recordings in which the dyadic subjects co-watched different positive-emotion video clips. These result in a total of 3600 time steps. Thus, the EEG input data for one subject has a shape of (30, 3600).

\begin{figure*}[t]
    \centering
    \includegraphics[width=1.0\linewidth]{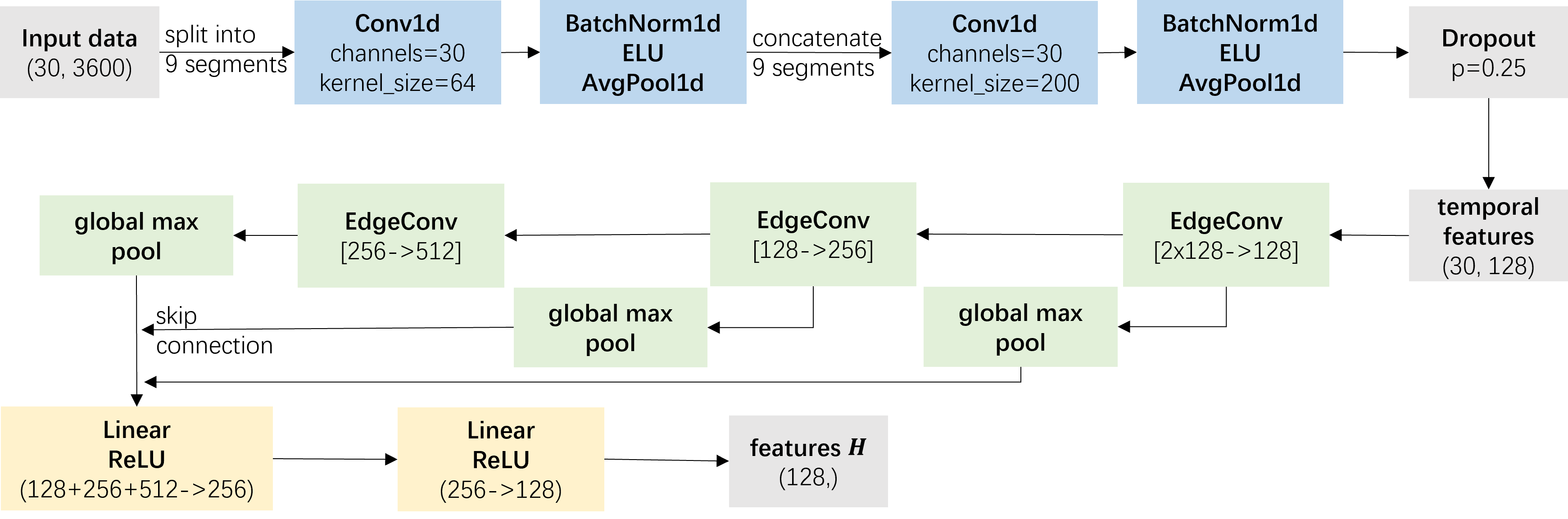}
    \caption{The detailed framework of the feature extractor in our proposed DSEN.}
    \label{fig:df}
\end{figure*}

Initially, for a dyadic subjects, we individually apply the feature extraction module of our DSEN to the EEG data of each subject. As illustrated in Fig. \ref{fig:df}, we first implement two 1D convolutional layers that accept inputs from 30 channels, and with a kernel size of 64 and 200, respectively. These layers extract temporal features while preserving information across electrode channels. For the first convolutional layer, the input data, corresponding to different positive-emotion video clips, is partitioned into 9 segments. We then individually extract local features from each segment. This is followed by batch normalization, an activation function, and pooling to a size of 100 for each of the local feature. After concatenating all the segments of local features, the second 1D convolutional layer is applied to extract global temporal features with a shape of (30, 128). 

We then, as described in Sec. 3.2 of our paper, take all the electrode channels as vertices in a fully-connected graph to feed a series of EdgeConv block that could aggregate and learn representations with the features from all neighboring vertices of each vertex. The initial EdgeConv block takes an input with twice the number of temporal features due to the concatenation of neighboring features. With two linear fully connection layers and a activation function (ReLU) in a EdgeConv block, the feature size is mapped to 128. The following two EdgeConvs have similar patterns that double the input features and finally to size of 512. Meanwhile, post feature extraction in each EdgeConv block, we employ a global max-pooling operation to aggregate vertex features. For further linear transformations, the pooled outputs from all three layers that contains feature size of 128, 256 and 512, respectively, are concatenated along the feature dimension. This operation fuses the graph-level representations obtained at different blocks, ensuring that the final representation contains multi-scale features. The subsequent two linear layers further refines the concatenated features to a dimensionality of 128.

\begin{figure*}[htbp!]
    \centering
    \includegraphics[width=0.8\linewidth]{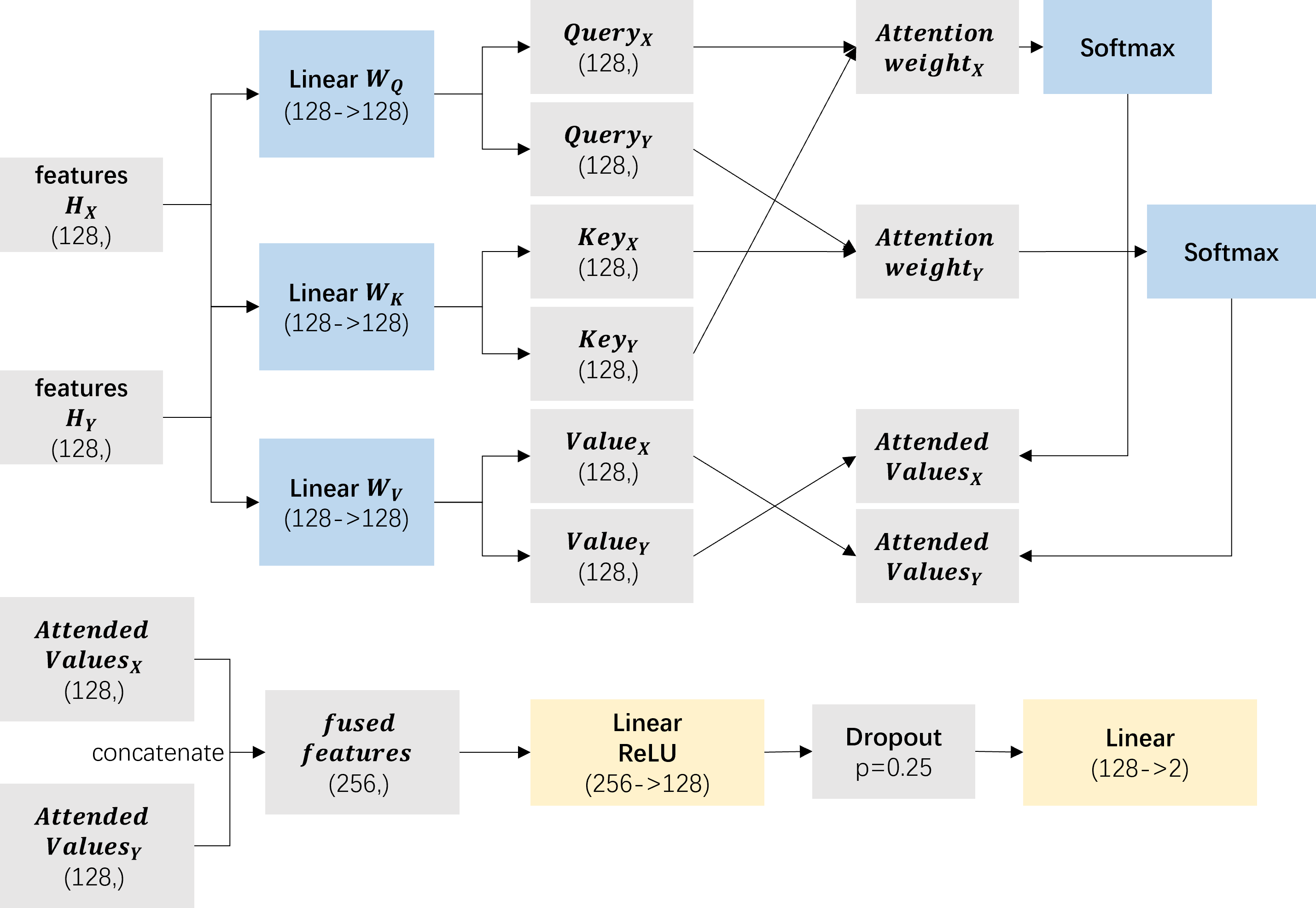}
    \caption{The detailed framework of the classifier in our proposed DSEN.}
    \label{fig:dc}
\end{figure*}

With the features of dyadic subjects (denoted as $X$ and $Y$) extracetd from aforementioned module, we then implement a module for social-relationship classification task. As illustrated in Fig. \ref{fig:dc} In this module, we first fused the dyadic-subject features utilizing an attention mechanism. We implement three linear learnable matrices $W_Q, W_K, W_V$ to calculate the Queries $Q$, Keys $K$ and Values $V$ for both $X,Y$. The $attention\ weight$ of $X$ is computed by taking the dot product of $Q_X$ with $K_Y$. Afterwards, the $attention\ weight$ undergoes a softmax operation and then multiplies with $V_Y$ to yield the $attended\ values$ for $X$. The same process is applied reciprocally for $Y$. By concatenating the $attended\ values$ of the two subjects, we obtain the fused features for the dyadic-subject EEG data. Subsequently, we transform these features through two linear layers. The first fully connected layer reduces the dimension of the fused features from 256 to 128, and the second one produces the classification result, aiming to predict the social relationship between the two subjects.

In our experiments, we tuned various hyperparameters to achieve optimal performance. To ensure reproducibility of our experiments and to provide a reference for researchers wishing to replicate our results, we detail all the hyperparameters employed in our experiments herein.  We set the \textbf{learning rate} at 0.0001 to ensure stability during the training process. The model was trained with a \textbf{batch size} of 79, striking a balance between computational efficiency and generalization. For our classification task, we use the \textbf{cross-entropy loss}, which is combined with the CCA loss in our proposed framework. The \textbf{maximum epoch number} is set to 100, preventing overfitting while allowing the model to converge. To further mitigate overfitting and introduce regularization, as highlighted in Fig. \ref{fig:df} and \ref{fig:dc}, we applied a dropout rate of 0.25. Finally, the \textbf{Adam optimizer} was utilized.

\subsection{Detailed Construction of Dyadic-Subject EEG Dataset}
\begin{figure}[htbp!]
    \centering
    \includegraphics[width=0.4\linewidth]{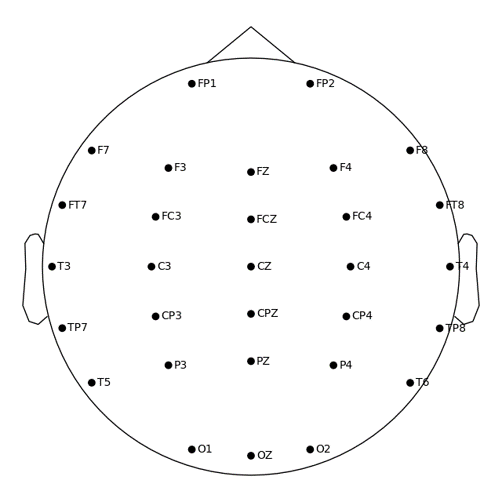}
    \caption{The topographical map of the EEG electrode positions in our experiments.}
    \label{fig:eegtopo}
\end{figure}

In this section, we provide a detailed introduction to the dyadic-subject EEG dataset we proposed, including the experimental procedures for data collection and examples from the dataset. As described in the Sec. 4.1 in our paper, EEG signals were recorded using the NuAmps 40-channel unipolar system with sampling frequency of 1000Hz. Two reference electrodes were attached to the mastoid areas on both sides of the subjects. The impedance of all electrodes was maintained below $5 k\Omega$. A low-pass filter was set to 60Hz. Furthermore, the sampling resolution is 22 bits and we employed 32 active silver/silver chloride (Ag/AgCl) electrodes for recording. The topographical map of the EEG electrode positions is depicted in Fig. \ref{fig:eegtopo}. 

\begin{figure} [htbp!]
    \centering
    \includegraphics[width=0.7\linewidth]{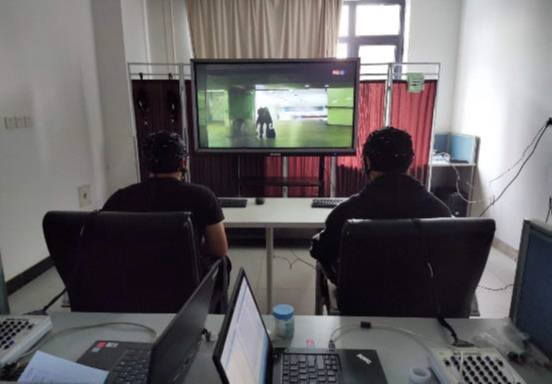}
    \caption{The scenario of our EEG hyper-scanning experimental task.}
    \label{fig:expscene}
\end{figure}

To provide a detailed description of our experimental scenario as an extension of our paper, we typically had two to three experimenters present to ensure the subjects had rested sufficiently and were in good health before assisting them in wearing our EEG recording devices. When capturing resting-state EEG data, we instructed the subjects to fixate on a cross located centrally on a screen (placed 1.5 meters away from their eyes). Subjects were then asked to remain seated quietly for a total of 8 minutes, split as 4 minutes with their eyes closed and another 4 minutes with eyes open. In the depicted hyper-scanning experiments, as shown in Fig. \ref{fig:expscene}, subjects were seated half a meter apart, co-watching video content on a shared screen, and not allowed to communicate with each other.

For the main experimental task, as mentioned in the our paper, we presented the dyadic subjects with nine video clips, each corresponding to a distinct \textit{positive emotional} label. To elaborate, these nine clips, selected from chinese emotional film database, include \textbf{``A Father's Journey"} for \textit{kinship}; \textbf{``Someone Secretly Loves You"} for \textit{gratitude}; \textbf{``Seeking Mr. Right (Anchoring in Seattle)"} for \textit{love}; \textbf{``How Insignificant We Are"} for \textit{awe}; \textbf{``Those Tempting Moments of Cuisine"} for \textit{longing}; \textbf{``The Little Refrigerator Runs Away from Home"} for \textit{friendship}; \textbf{``Detective Chinatown II"} for \textit{humor}; \textbf{``China's Gold-winning Moments in the Rio Olympics"} for \textit{pride}; and \textbf{``When Someone Asks About Qian Xuesen"} for \textit{respect}. 

In our novel proposed dyadic-subject EEG dataset, a total of 72 pairs of subjects are included. Among them, 18 pairs are strangers to each other, while the remaining 54 pairs are acquaintances. As previously detailed, we recorded the EEG data of each subject as they co-watching nine video clips. By segmenting the EEG recordings using a 2-second sliding window, we were able to obtain 5688 data samples. Each of these samples contains the EEG data from dyadic subjects. Structurally, a single data sample is organized as $2 \times 30 \times 3600$, representing readings from 30 electrode channels over 3600 time steps of 2 subjects. 

\subsection{Comparison across Different Number of Video Clips}
\begin{figure}[htbp!]
    \centering
    \includegraphics[width=1.0\linewidth]{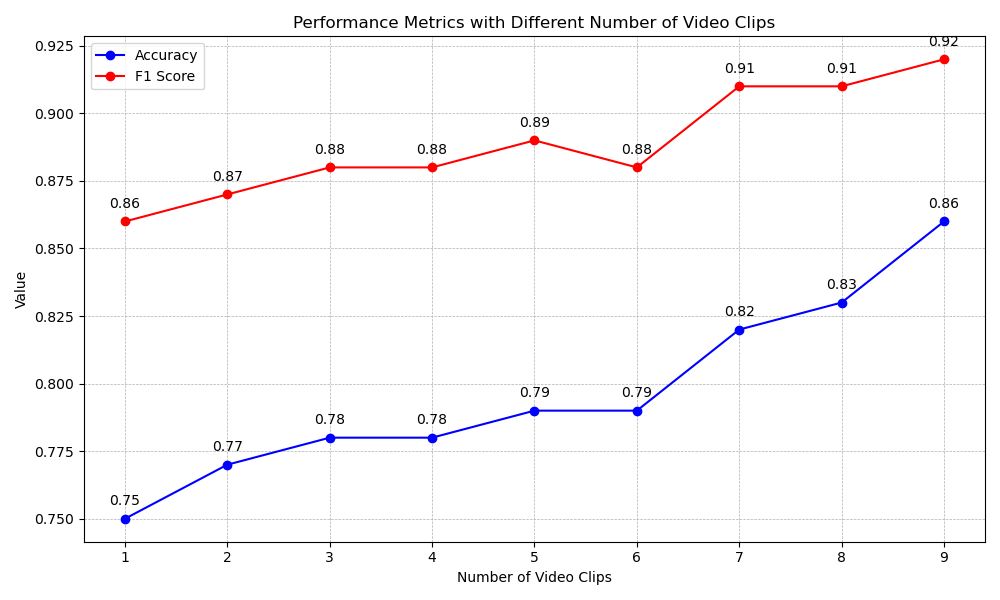}
    \caption{The result of the experiment that compares the input dyadic EEG data of the subjects co-watching different numbers of video clips.}
    \label{fig:diffnumvideos}
\end{figure}

To further investigate whether we need EEG data from the dyadic subjects co-watching all nine video clips to classify the social relationship between them, we conducted an additional experiment which is training our model with the dyadic EEG data from the subjects co-watching different number of video clips. In this experiment, we sequentially remove EEG segment corresponding to one random video clips from our data samples and continued this removal process until we are left with EEG data from just one segment for relationship prediction. The averaging accuracy and F1 score results across multiple experiments (nine in this experiment) are illustrated in Fig. \ref{fig:diffnumvideos}. Notably, In our proposed feature extractor, the second 1-D convolution layer employed a large kernel size, designed to capture global-temporal patterns. However, when adapting to EEG data derived from subjects watching just a single video clip, the total time steps in that sample is 100 which is less than our constant kernel size. Consequently, we adjusted the kernel size down to 100 to fit the EEG data associate with 1 video clip. The experimental results indicate that, despite minor fluctuations, as the number of video clips co-watched by two subjects increases, each dyadic EEG sample could contain more diverse information, enhancing the accuracy of discriminating their relationship. This aligns with our intuitive expectations, as richer emotional perspective can reveal more inter-subject dynamics and patterns.

\subsection{T-test for Gender of two participants}
\begin{table}[t]
    \centering
    \begin{tabular}{l|l|c|c}
    \hline
    IV & feature & statistic & p-value \\
    \hline
    \multirow{8}{*}{genders} & $PLV_{\theta}$ & -1.019 & 0.312 \\
     & $PLV_{\alpha}$ & 0.581 & 0.563 \\
     & $PLV_{\beta}$ & 1.055 & 0.295 \\
     & $PLV_{\gamma}$ & 0.628 & 0.532 \\
     & $ISC_{\theta}$ & -0.400 & 0.690 \\
     & $ISC_{\alpha}$ & -0.616 & 0.540 \\
     & $ISC_{\beta}$ & 0.621 & 0.537 \\
     & $ISC_{\gamma}$ & 1.471 & 0.146 \\ \hline
    \end{tabular}
    \caption{t-test results with $ISC$ and $PLV$ from dyadic-subject EEG data as dependent variables and gender of them as independent variables.}
    \label{tab:ttest-gender}
\end{table}
The significance level was set to 0.05. Specifically, the gender composition of the two subjects was categorized into same gender and different genders. Table \ref{tab:ttest-gender} demonstrates that the EEG synchrony features $ISC$ and $PLV$ across all frequency bands for dyadic subjects show no significant differences across different gender types. The experimental results lead us to believe that gender does not have a significant impact on differentiating EEG data between various dyadic subjects. Therefore, we can disregard the influence of varying gender compositions among subjects in the experiments.

\end{document}